\documentclass{aastex701}
\usepackage{amsmath}

\begin{document}

\title{Null Results, Real Learning: Geomagnetic Response to an X1.8 Solar Flare with Research-Grade and Smartphone Magnetometers in a Citizen-Science Classroom Activity}

\author[orcid=0000-0002-2945-5530,gname=Roger,sname=Hart]{Roger M. Hart}
\affiliation{Community College of Rhode Island, Department of Physics and Engineering, Flanagan Campus, 1762 Old Louisquisset Pike, Lincoln, RI 02865}
\email[show]{Rmhart1@ccri.edu}

\author[gname=Lauren,sname=Messina]{Lauren E. Messina}
\affiliation{Community College of Rhode Island, Department of Physics and Engineering, Flanagan Campus, 1762 Old Louisquisset Pike, Lincoln, RI 02865}
\email[show]{lemessina@ccri.edu}

\author[gname=Eric,sname=Schenck]{Eric A. Schenck}
\affiliation{Community College of Rhode Island, Department of Physics and Engineering, Flanagan Campus, 1762 Old Louisquisset Pike, Lincoln, RI 02865}
\email{lemessina@ccri.edu}

\author[gname=Samantha,sname=Kaplan]{Samantha R. Kaplan}
\affiliation{Community College of Rhode Island, Department of Physics and Engineering, Flanagan Campus, 1762 Old Louisquisset Pike, Lincoln, RI 02865}
\email{lemessina@ccri.edu}

\author[gname=Diego,sname=Cant\'{e}]{Diego A. Cant\'{e}}
\affiliation{Community College of Rhode Island, Department of Physics and Engineering, Flanagan Campus, 1762 Old Louisquisset Pike, Lincoln, RI 02865}
\email{lemessina@ccri.edu}

\author[gname=Izaiah,sname=Figueroa]{Izaiah Figueroa}
\affiliation{Community College of Rhode Island, Department of Physics and Engineering, Flanagan Campus, 1762 Old Louisquisset Pike, Lincoln, RI 02865}
\email{lemessina@ccri.edu}

\author[gname=Gabriella,sname=Sepe]{Gabriella Sepe}
\affiliation{Community College of Rhode Island, Department of Physics and Engineering, Flanagan Campus, 1762 Old Louisquisset Pike, Lincoln, RI 02865}
\email{lemessina@ccri.edu}

\author[gname=Zavier,sname=Lopez]{Zavier Lopez}
\affiliation{Community College of Rhode Island, Department of Physics and Engineering, Flanagan Campus, 1762 Old Louisquisset Pike, Lincoln, RI 02865}
\email{lemessina@ccri.edu}

\author[gname=Ryan,sname=Ward]{Ryan Ward}
\affiliation{Community College of Rhode Island, Department of Physics and Engineering, Flanagan Campus, 1762 Old Louisquisset Pike, Lincoln, RI 02865}
\email{lemessina@ccri.edu}

\author[gname=Sammy,sname=Morse]{Sammy P. Morse}
\affiliation{Community College of Rhode Island, Department of Physics and Engineering, Flanagan Campus, 1762 Old Louisquisset Pike, Lincoln, RI 02865}
\email{lemessina@ccri.edu}

\author[gname=Melanie,sname=Ramirez]{Melanie V. Ramirez}
\affiliation{Community College of Rhode Island, Department of Physics and Engineering, Flanagan Campus, 1762 Old Louisquisset Pike, Lincoln, RI 02865}
\email{lemessina@ccri.edu}

\author[gname=Brady,sname=Gaulin]{Brady J. Gaulin}
\affiliation{Community College of Rhode Island, Department of Physics and Engineering, Flanagan Campus, 1762 Old Louisquisset Pike, Lincoln, RI 02865}
\email{lemessina@ccri.edu}

\begin{abstract}

Introductory college courses in Earth and space science offer rich opportunities for citizen-science projects. One especially compelling context is Earth’s geomagnetic field: a self-excited fluid dynamo in the planet’s electrically conducting outer core generates a global magnetic field that couples Earth’s interior to solar forcing via the magnetosphere, and the resulting geomagnetic variability provides a natural laboratory for space-weather education and citizen-science engagement. We examined the viability of smartphone magnetometers for quantitative geomagnetic monitoring during the 4~November~2025 X1.8 solar flare, linking planetary magnetism, space weather, and authentic undergraduate research. Co-located observations were obtained using a Geometrics~G-857 proton-precession magnetometer and tri-axial smartphone magnetometers logging through \textit{Physics Toolbox}. The field campaign, embedded in a course-based undergraduate research experience (CURE), emphasized the Nature of Science (NOS) through student-led data acquisition, synchronization, and comparison with INTERMAGNET and GOES--18 data products. Statistical analysis of fourteen one-minute paired averages spanning 17{:}27--17{:}40~UT revealed a systematic smartphone bias of $+626.3\pm80$~nT (95\%~CI~546--707~nT) relative to the G-857, with a weak negative correlation ($r\approx-0.38$). These results confirm that smartphone magnetometers lack the precision and calibration stability required for detecting nanotesla-scale flare signatures, though they remain valuable as pedagogical and engagement tools. Framed within a tiered ``instrumentation ladder,'' we propose an integrated hierarchy linking research-grade observatories, intermediate-cost community magnetometers (e.g., HamSCI Personal Space Weather Stations), and smartphones as high-engagement entry points for participation in geomagnetic and space-weather studies. This structure aligns citizen science with open-data protocols and NOS pedagogy, transforming low-cost sensing from novelty into epistemically grounded inquiry that is feasible for introductory college laboratories. Under realistic field conditions, smartphones provide conceptual access and metacognitive insight but not quantitative resolution; a hybrid network---combining calibrated instruments with distributed citizen participation---offers a scalable, educationally rich approach to mapping Earth’s magnetic environment and understanding its dynamic solar coupling.
\end{abstract}



\section{Introduction}
Planetary magnetic fields are macroscopic manifestations of internal fluid dynamos operating in electrically conducting regions of planets; in our Solar System, bodies with substantial convecting conductive interiors---such as Earth’s liquid iron outer core or the metallic hydrogen layers of Jupiter and Saturn---can sustain self-excited dynamo action that generates large-scale, approximately dipolar fields (e.g., de Pater \& Lissauer, 2015; Stevenson, 2010). These fields carve out magnetospheres from the solar wind, guiding charged particles and producing phenomena such as aurorae and radiation belts, and underpinning the broader framework of magnetospheric physics developed from \emph{in situ} measurements and theory (e.g., Kivelson \& Russel, 1995; Kivelson et al., 1996; Van Allen, 2004; de Pater \& Lissauer, 2015). 

The strength and morphology of a planet’s field are controlled by interior structure, convective power, and rotation rate, yielding a diversity of magnetic characteristics across the Solar System: Earth’s field is nearly dipolar and globally enveloping, Uranus and Neptune host highly tilted, offset multipolar fields, Mercury maintains a weak but active dynamo, and Mars and the Moon lack global fields yet retain remnant crustal magnetization that encodes ancient dynamo activity (See further de Pater \& Lissauer, 2015; Stevenson, 2010). These contrasts can be interpreted in terms of variations in core size, cooling history, and core phase changes (e.g., inner-core nucleation or precipitation of light elements), so that magnetic observables provide a window into planetary thermochemical evolution (Stevenson, 2010). Within this comparative framework, Earth’s geomagnetic field remains the best-resolved planetary dynamo and a critical reference point: it is described using global spherical-harmonic models that separate internal (core) and external (ionospheric, magnetospheric) sources and track secular variation, geomagnetic jerks, and polarity reversals, combining observatory networks such as USGS and INTERMAGNET with high-precision satellite missions and models like the World Magnetic Model (Kono, 2007; Alken et al., 2021). Together, these observations demonstrate how a single planetary dynamo can be studied continuously from seconds-long space-weather disturbances to million-year-scale polarity reversals, linking interior dynamics to magnetospheric structure and heliospheric coupling (Kono, 2007; Van Allen, 2004).

\subsection{Smartphones and Low-Cost Sensing in Science Education}
Over the last decade, smartphones and low-cost sensors have become routine geoscience tools. Their built-in sensors turn them into ``labs in your pocket'' (Vieyra \& Vieyra, 2019). Apps like Physics Toolbox log multi-sensor data for class, home, and field (Vieyra \& Vieyra, 2019). Thus smartphones support physics and geoscience labs from kinematics to electromagnetism. Prior work showed ride and ``kitchen physics'' accelerations and rotations are well captured (Vieyra \& Vieyra, 2014; Vieyra, Vieyra, \& Macchia, 2017). These activities foreground calibration, error analysis, and experimental design through student-collected data.

Educators then built structured smartphone challenges and curricula. ``Turn Your Smartphone Into a Science Laboratory'' offers five NGSS-aligned sensor challenges (Vieyra et al., 2015). Such tasks let phones stand in for probeware, broadening hands-on access. Gamified tools like Physics Toolbox Play teach physics via sensor-based games (Vieyra et al., 2020). Together, these efforts cast smartphones as a ``digital science education revolution'' platform (Vieyra, 2018; Vieyra et al., 2023). Class sets of phones can form distributed sensor networks for collaborative experiments (e.g., Kim et al., 2024).

New work adds AR and embodied learning to make hidden phenomena visible. LiDAR-enabled devices power ``Motion Visualizer'' position--time graphing from live motion. Students move and instantly see their motion graphs, linking kinesthetic feel to slopes (Megowan-Romanowicz et al., 2023; O’Brien et al., 2023). Studies report better motion-graph interpretation with these AR LiDAR tools (Megowan-Romanowicz et al., 2023; Vieyra et al., 2024). Magna-AR instead overlays 3D magnetic field lines using the phone magnetometer (Vieyra \& Vieyra, 2022). Learners scan coils or bar magnets, building concrete 3D mental models of fields. Other apps let students reshape virtual magnets and see fields, energy, and space-weather links (Vieyra et al., 2024). Overall, smartphones act as both sensors and interactive visualization portals.

Still, smartphones are consumer devices whose limits matter in serious use. Noise, orientation, dynamic range, and nearby metal can degrade phone measurements. Yet controlled tests show smartphone sensors can perform surprisingly well. One study found reliable dynamic-balance indices from phone accelerometers on a board (Atalay \& Turan Kızıldoğan, 2025). Likewise, low-cost geophysical rigs can mimic commercial resistivity and refraction tools (e.g., Clark \& Page, 2011). These systems demand more manual work and calibration but still yield ``good enough'' data. The necessary trade-offs become lessons in noise, effort, and real-world data challenges. Thus limitations themselves become teachable moments about measurement, calibration, and interpretation (Vieyra \& Vieyra, 2014, 2019; Clark \& Page, 2011; Vieyra et al., 2015, 2020, 2024). These concepts are central to general-education Earth and Space Science classes taught at both the high school level (e.g., NGSS Lead States, 2013) and introductory college classes (e.g., Fraknoi et al., 2022). 

\subsection{Citizen Science, Public Engagement, and the Nature of Science}
The proliferation of accessible sensing technologies dovetails with a mature citizen-science landscape in which volunteers contribute to environmental and physical sciences. Frameworks distinguish contributory projects (scientist-designed, with volunteers primarily collecting or processing data), collaborative projects (with volunteer input on design, analysis, or dissemination), and co-created projects (in which community members and scientists jointly define questions and protocols) (Bonney et al., 2016). Large-scale contributory efforts such as biodiversity networks and environmental monitoring campaigns routinely produce peer-reviewed datasets, showing that volunteer-collected data, when carefully managed, can meet or approach professional standards (Fraisl et al., 2022; Frigerio et al., 2021). In astronomy, projects like Galaxy Zoo demonstrate that non-experts can classify galaxies, discover exoplanets, and identify previously unrecognized phenomena via online platforms, sometimes prompting research questions unanticipated by professional teams (Marshall et al., 2015). Emerging work in geospace and atmospheric science similarly envisions citizen sensor networks---from auroral cameras and airglow imagers to personal GNSS receivers and backyard magnetometers---that contribute direct measurements of the near-Earth space environment and augment official monitoring, especially during rare or extreme events (Grandin et al., 2025).

Yet producing new data or publications through citizen science does not automatically yield substantial learning gains by students. Reviews show that improvements in understanding scientific content and processes are often modest or highly variable unless they are explicitly targeted in project design (e.g., Bonney et al., 2016). In many contributory projects, volunteers classify or collect data without knowing why it is needed, limiting conceptual gains. Authors therefore advocate explicitly reflective designs---training that explains why protocols are structured as they are and prompts that ask volunteers to connect observations to hypotheses (Bonney et al., 2016; Fraisl et al., 2022). These recommendations parallel formal science-education work on the Nature of Science (NOS): learners often hold alternative conceptions or views, and approaches that expect authentic NOS conceptions to emerge implicitly from activities or labs have generally failed (See Abd-El-Khalick, 2001). Instead, explicit reflection on how scientific knowledge is generated, coupled with NOS themes embedded directly in content courses, could produce more robust shifts, though even future teachers struggle to implement this consistently (Abd-El-Khalick, 2001).

One response is to design data-rich, societally relevant modules that intertwine content learning, NOS, and citizen-science-like practices. For example, Measuring the Earth with GPS has students analyze real GPS time series to investigate plate motions, crustal deformation, and climate-related ice mass changes, while foregrounding how geoscientists collect, interpret, and evaluate data as evidence (Kortz \& Smay, 2019). Evaluations indicate gains in scientific argumentation and appreciation of how knowledge emerges from observations (Kortz \& Smay, 2019). Place-based (e.g., Powers, 2004) and service-learning (e.g., Lima, 2023; Strait \& Lima, 2023) approaches can be applied in community college astronomy courses---through skyglow surveys, public observing events, or student-designed projects on campus environmental issues---yielding improved conceptual understanding, motivation, and scientific identity (e.g., Hart et al., 2025). 

Data quality and open-science practices weave through these efforts: stakeholders value different aspects of ``quality,'' and best practices emphasize clear protocols, training, calibration, verification, and transparent reporting of limitations and error controls (Balázs et al., 2021; Fraisl et al., 2022). When participants grapple with uncertainty, noise, and bias---comparing citizen-generated data with benchmarks, interrogating outliers, or learning how smartphone and low-cost sensors require attention to orientation, thermal effects, and other artifacts---they are effectively practicing calibration, noise filtering, and validation, skills central to sophisticated NOS conceptions. Thus, projects that pair rigorous data-quality procedures with explicit NOS reflection are positioned to deliver both credible scientific products and meaningful learning outcomes (Bonney et al., 2016; Frigerio et al., 2021).

\subsection{Solar Flares and Geomagnetic Disturbances as a Natural Laboratory}
Against this backdrop of planetary magnetism, accessible sensing, and citizen engagement, we focus on geomagnetic disturbances driven by solar flares, primarily because of how engaging the effects of geomagnetism can be via aurora to the general public (e.g., Johannesson et al., 2017). Powerful X-class flares emit bursts of ionizing X-ray and EUV radiation and may launch CMEs that perturb Earth’s space environment. On the dayside, enhanced ionization rapidly strengthens ionospheric currents, producing ground magnetic perturbations historically termed solar flare effects (SFE) or ``magnetic crochets.'' Classical studies show that SFE amplitude and polarity depend on flare intensity and spectrum, local time, solar zenith angle, latitude, and the pre-existing state of ionospheric currents, so that even large flares can generate clear signatures at some observatories and barely detectable signals at others (e.g., Curto et al., 1994, 2009, 2020; Hart, 2025). Because only the largest flares yield unambiguous SFEs, they provide episodic but valuable ``impulses'' to the magnetosphere--ionosphere system, allowing researchers to infer properties of ionospheric current systems (Curto et al., 2020) and to assess operational risks from impulsive geomagnetic disturbances that arrive with little warning (e.g., Curto et al., 2020; Sokolova et al., 2021; Vieyra \& Lopez, 2023; Evans et al., 2024; Griffin et al., 2025).

For educators and citizen-science organizers, such flares are attractive natural experiments: they unfold on human timescales (seconds--minutes), affect much of the dayside, and connect solar physics, geomagnetism, and technological impacts (e.g., Sokolova et al., 2021; Vieyra \& Lopez, 2023; Evans et al., 2024; Griffin et al., 2025). An anticipated or promptly identified flare can trigger a coordinated campaign in which students and amateur observers check their magnetometers---research-grade or smartphone-based---at a common universal time. Participants must synchronize clocks, account for location and sensor orientation, compare records across sites, and relate their measurements to satellite X-ray flux, practicing authentic scientific skills. Both detections and non-detections become instructive: differences between sites can be traced to latitude, local noise, or instrument configuration, prompting discussion of ionospheric current geometry, signal-to-noise, and instrument sensitivity. Thus, a flare campaign can integrate content on Earth’s magnetosphere and space weather with process skills and Nature-of-Science (NOS) reflections on data and instrumentation.

We use the 4 November 2025 X1.8 flare as a testbed for two linked questions, one scientific and one educational. Scientifically, we test whether smartphone magnetometers can track co-located measurements from a research-grade proton magnetometer and nearby observatories during a space-weather event closely enough to support meaningful geomagnetic interpretation, and what calibration and processing are needed to make their data usable. Educationally, we embed this flare-focused, multi-instrument campaign in an undergraduate course as a citizen-science project, asking what students learn about geomagnetism and the Nature of Science (NOS) by acting as co-investigators in authentic data collection and analysis. By situating the work at the intersection of planetary magnetism, citizen science, and education research, we (1) quantify smartphone magnetometer sensitivity, accuracy, and noise relative to a professional instrument during an X-class flare, (2) present a transferable model for course-based citizen science that integrates NOS instruction into disciplinary content, and (3) outline a pathway from single-class projects to regional school-based smartphone magnetometer networks that expand spatial coverage while engaging many learners. In doing so, we show how Earth’s magnetic fluctuations---measured even with everyday smartphones---can serve as a powerful vehicle for public engagement and learning, exemplifying a productive convergence of scientific and educational aims aligned with broader movements toward participatory, inclusive science.

\section{Materials and Methods}
\subsection{Overview}
This study compared co-located magnetic-field measurements from a Geometrics G-857 proton magnetometer (Hart et al., 2023; Hart, 2024; Hart et al., 2025) with those from tri-axial smartphone magnetometers during the interval surrounding the 4 November 2025 X1.8 solar flare. The objective was to determine whether smartphone-derived total-field magnitudes differ significantly from research-grade absolute total-field measurements under real-time space-weather conditions at mid-latitudes. We provide a detailed, complete, and reproducible workflow for data acquisition, synchronization, quality control, scalar-field computation, temporal resampling, and cross-instrument statistical comparison. This could be adapted to any community college setting, especially with the development of lower cost magnetometers (e.g., Kim et al., 2024). 

\begin{figure}
  \centering
  \includegraphics[width=\textwidth]{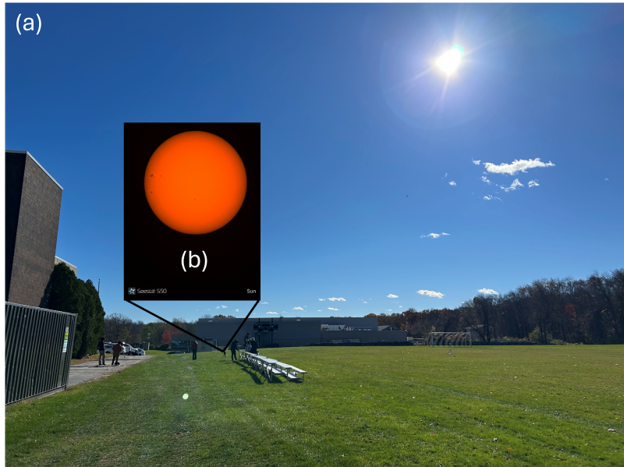}
  \caption{(a) Collaborative observations of multi-parameter data collection of the atmosphere, geomagnetic field, and the Sun. Observations completed, while not a part of this study, included meteorological conditions (wind speed, humidity, temperature, barometric pressure), generalized ionizing radiation using a Geiger counter, and sunspot measurements using multiple telescopes, chiefly two Seestar S50 all-in-one smart telescopes and a Celestron SkyProdigy 6 computerized telescope, all with solar filters. (b) Example astrophotograph of the Sun using default Seestar S50 specifications around the time of the 4 November 2025 X1.8 solar flare.}
  \label{fig:campaign}
\end{figure}

\subsection{Instructional Framing and Collaboration}
This project was designed as a joint science-and-education experiment embedded in an introductory laboratory sequence in a course after Kortz et al. (2025). Based on Course-Based Undergraduate Research Experiences (CUREs) which have indicated that undergraduate research is widely recognized as an effective way to support science learning, but there are still relatively few detailed published examples designed for introductory-level students (e.g., Kortz \& Van Der Hoeven Kraft, 2016). The course instructor and students led the design of the observing campaign, the configuration of instruments, data curation, the processing pipeline, and manuscript preparation, with help from the laboratory technician. Students contributed as collaborators in data handling, time alignment, and scalar-magnitude comparisons at keyed times around the flare peak (17{:}34~UT).

The analytical workload for students was intentionally constrained to reading off total-field values and computing simple magnitude differences ($\Delta F$), rather than attempting formal identification of magnetic crochets. This design aligned with course learning outcomes in quantitative reasoning and data literacy rather than advanced ionospheric diagnostics.

\subsection{Science Question and Event Selection}
The core science question guiding this study was how the local geomagnetic-field magnitude behaved near the 4 November 2025 X1.8 solar flare peak, and how campus values compared with regional observatories at mid-latitudes.

We targeted the X1.8 flare that peaked at 17{:}34~UT (12{:}34~EST), as identified in the GOES-18 0.1--0.8~nm soft X-ray flux. X-class flares can, under favorable conditions, produce short-lived dayside total-field perturbations via rapid changes in ionospheric conductance, sometimes manifesting as ``magnetic crochet'' signatures at ground stations. The occurrence, polarity, and amplitude of such signatures depend strongly on local time, solar zenith angle, latitude, pre-existing geomagnetic activity, and the detailed ionospheric conductivity profile (Curto et al., 1994, 2009; Curto, 2020). This event therefore provided a realistic but non-ideal test of whether consumer-grade smartphone magnetometers could track absolute field magnitude and any flare-time perturbations in parallel with a research-grade proton magnetometer.

\begin{figure}
  \centering
  \includegraphics[width=\textwidth]{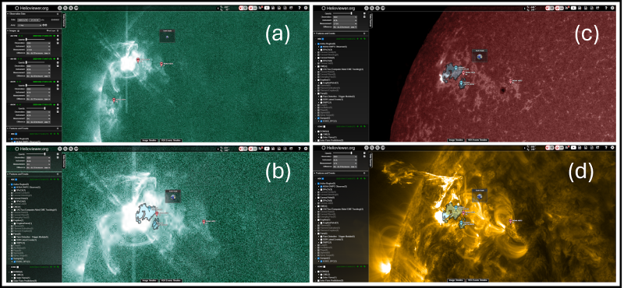}
  \caption{Multispectral images of the Sun taken by the Solar Dynamics Observatory (SDO) using the Atmospheric Imaging Assembly (AIA) instrument, accessed through the Helioviewer Project. Panels (a) and (b) show a zoomed view of the flaring active region in a hot 94~\AA{} band, emphasizing the compact flare core and overlying coronal loops; panel (c) shows the same region in a 1700~\AA{} band that highlights the underlying sunspots and chromospheric structure; and panel (d) reveals extended flare structures and surrounding arcade in 175~\AA{}. The bright arch-like features visible in panel (d) are coronal loops---distinct, arch-shaped structures of relatively dense plasma confined within magnetic flux tubes, anchored at two footpoints in the photosphere and extending upward through the transition region into the lower corona. In each panel, overlays and markers indicate the region of interest and sampling locations used in our analysis of the 4 November 2025 X1.8 flare.}
  \label{fig:sdo}
\end{figure}

\subsection{Observational Campaign and Context Measurements}
The flare was observed as part of a broader, multi-parameter ``Sun--Earth'' observing campaign on campus. In addition to geomagnetic measurements, student teams recorded meteorological variables such as wind speed, humidity, temperature, and barometric pressure using handheld weather stations; generalized ionizing radiation levels using a Geiger counter; and solar photospheric imagery and sunspot morphology using multiple telescopes equipped with solar filters, specifically two Seestar S50 all-in-one smart telescopes and a Celestron SkyProdigy 6 computerized telescope. These contextual measurements are not analyzed quantitatively in this paper, but they provide an authentic, multi-sensor framework within which the magnetometry activities were embedded and support future cross-variable analyses of space-weather events.

\subsection{Instrumentation}
The primary geomagnetic instrument was a Geometrics G-857 proton precession magnetometer, operated as an absolute scalar magnetometer measuring the total field magnitude, $F(t)$, in nanotesla (nT). The G-857 provides absolute measurements with high long-term stability and negligible dependence on sensor orientation. For this campaign, data were logged at approximately 10-second cadence, with each record including date, time, total-field magnitude $F$, internal temperature, and an instrument quality flag. The sensor head was mounted outdoors on a rigid, non-magnetic platform, deliberately sited away from buildings, vehicles, and obvious sources of cultural magnetic noise. The G-857 total-field series serves as the reference ``truth'' dataset for local geomagnetic conditions.

Multiple student smartphones were deployed as auxiliary magnetometers. Each device ran Physics Toolbox Sensor Suite, which exposes the built-in tri-axial magnetometer and logs magnetic-field components $B_x$, $B_y$, and $B_z$ in microtesla ($\mu$T), together with operating-system timestamps (Vieyra \& Vieyra, 2019). Smartphones were placed near the G-857 sensor head in a fixed orientation and held away from metal stands, cables, and other ferromagnetic materials to minimize local bias fields and orientation artifacts. Sampling occurred at sub-second cadence, providing high-frequency, multi-axis measurements suitable for post hoc scalar-field computation and statistical comparison.

Smartphone magnetometers are known to exhibit higher noise levels, orientation dependence, and local bias relative to absolute proton magnetometers. However, they offer accessible, low-cost platforms that have been widely used for physics and geoscience education (Vieyra \& Vieyra, 2014; Vieyra et al., 2015; Vieyra, Vieyra, \& Macchia, 2017; Megowan-Romanowicz et al., 2023; O’Brien et al., 2023; Vieyra et al., 2024), and under constrained procedures can yield robust measurements for some indices (Atalay \& Turan Kızıldoğan, 2025).

To contextualize campus observations, we retrieved USGS/INTERMAGNET one-minute total-field data from nearby mid-latitude observatories, providing independent regional records of $F(t)$. We also downloaded the GOES-18 soft X-ray 0.1--0.8~nm flux curve from 4 November 2025 to define the flare onset, peak, and decay phases (Benz, 2017; Curto, 2020). These external datasets served two roles: they allowed benchmarking of the campus proton magnetometer against well-characterized observatories, and they anchored the timing of student tasks to a clearly defined solar driver.

\subsection{Time Window and Synchronization}
We adopted a flare-centered analysis window spanning 17{:}27--17{:}41~UT, comfortably bracketing the 17{:}34~UT flare peak and allowing pre- and post-peak comparisons. For student activities and educational purposes, a broader $\pm$30-minute context window around the peak was also considered.

All timestamps were handled in Coordinated Universal Time (UTC). Smartphone timestamps were read directly from device logs and converted to UTC where necessary. G-857 timestamps were reconstructed by combining instrument date and time fields into full UTC datetime objects. After parsing, all timestamps were treated as naive UTC for export and analysis, preserving their universal temporal alignment. We verified the absence of clock discontinuities or visible drift between the G-857 and smartphone clocks over the 14-minute core window by comparing timestamp sequences and cross-checking against the known timing of the flare in the GOES-18 record.

\subsection{Computation of Smartphone Total-Field Magnitude}
Smartphone logs provided the tri-axial components $B_x$, $B_y$, and $B_z$ in $\mu$T. To compare with the scalar G-857 series, we computed the total-field magnitude at each smartphone timestamp. Magnetic-field components originally in microtesla were converted to nanotesla by multiplying each component by 1000:
\begin{equation}
  B_x^{(\mathrm{nT})} = 1000\, B_x^{(\mu\mathrm{T})}, \quad
  B_y^{(\mathrm{nT})} = 1000\, B_y^{(\mu\mathrm{T})}, \quad
  B_z^{(\mathrm{nT})} = 1000\, B_z^{(\mu\mathrm{T})}.
\end{equation}
The total magnetic-field magnitude was then computed at each time step from the three components:
\begin{equation}
  \lvert \mathbf{B} \rvert =
  \sqrt{ \left(B_x^{(\mathrm{nT})}\right)^2
       + \left(B_y^{(\mathrm{nT})}\right)^2
       + \left(B_z^{(\mathrm{nT})}\right)^2 }.
\end{equation}
This procedure produced a scalar time series $\lvert \mathbf{B}(t) \rvert$ in nT, directly comparable in units to the G-857's $F(t)$ measurements.

\subsection{Temporal Averaging and Resampling}
To mitigate high-frequency jitter, orientation sensitivity, and sensor noise in the smartphone data while preserving any plausible flare-time geomagnetic signature, we applied a two-stage averaging strategy. In the first stage, raw smartphone $\lvert \mathbf{B} \rvert$ values were averaged into 10-second bins to reduce sub-second noise and micro-variations associated with small hand or device motions. The G-857 data, already logged at approximately 10-second cadence, were retained at their native resolution.

In the second stage, both the smartphone and G-857 time series were resampled into one-minute bins for cross-instrument comparison. For each one-minute interval, all 10-second smartphone $\lvert \mathbf{B} \rvert$ values falling within that minute were averaged to produce a per-minute scalar field, and all G-857 total-field values within the same minute were also averaged. Minutes that lacked valid data for a given series were left as missing values; no interpolation was performed. This one-minute cadence approximates the temporal resolution of regional observatories and provides statistically stable samples for paired comparison.

\subsection{Data Quality Control and Statistical Analysis}
Before averaging and merging, we applied several quality-control steps. Records with non-parsable timestamps or inconsistent date--time fields were removed. Obvious spikes associated with app restarts, sensor glitches, or G-857 records flagged as invalid by the instrument quality flag were also removed. All measurements were standardized to nanotesla before averaging. The 10-second series were visually inspected to confirm that no large step changes or clock discontinuities were introduced by parsing or resampling.

No detrending, baseline removal, or high-pass or low-pass filtering was applied to any series. Because the primary goal was to compare absolute magnitudes between instruments and to support magnitude-only student tasks, rather than to detect subtle ionospheric crochets of order a few nanotesla, retaining the native trend structure was essential.

To make the analysis tractable for a first- or second-year audience while preserving scientific authenticity, we prepared time-aligned CSV files and plots that contained the campus G-857 total-field $F(t)$, the nearest USGS/INTERMAGNET observatory total-field $F(t)$, and the GOES-18 soft X-ray flux time series. Within the flare-centered $\pm$30-minute window, students were asked to read off $F(t)$ at keyed times, such as 17{:}30, 17{:}34, and 17{:}40~UT, from both the campus and nearest USGS observatory time series. They then computed simple magnitude changes, $\Delta F = F(T_2) - F(T_0)$, between pre-peak and peak or post-peak times for both locations. Finally, they wrote short comparative notes describing the direction (increase or decrease) and approximate magnitude of changes they observed, without attempting to classify features as magnetic crochets.

Smartphones were used as auxiliary instruments. Students collected short contextual time series using their smartphones, then compared rough smartphone magnitudes and qualitative trends with the G-857 and USGS data. This use of phones as ``labs in their pockets'' builds on a substantial literature demonstrating the pedagogical value of smartphone sensors for embodied, augmented-reality--mediated, and inquiry-based learning in physics and Earth--space science (Vieyra \& Vieyra, 2014; Vieyra et al., 2015; Vieyra, Vieyra, \& Macchia, 2017; Vieyra \& Vieyra, 2019; Megowan-Romanowicz et al., 2023; O’Brien et al., 2023; Vieyra et al., 2024; Atalay \& Turan Kızıldoğan, 2025).

For the research analysis presented here, the per-minute smartphone and G-857 series were merged on their common minute timestamps. For each minute with valid values from both instruments, we computed the difference
\begin{equation}
  \Delta_{\mathrm{1min}} = B_{\mathrm{phone,1min}} - F_{\mathrm{G857,1min}},
\end{equation}
which we interpreted as the smartphone--proton magnetometer offset at one-minute cadence. The resulting paired samples were used to address two questions: whether smartphone total-field magnitudes are statistically consistent with absolute proton-precession measurements, and whether minute-scale variations in the smartphone record track genuine geomagnetic variations captured by the G-857.

Magnetic-field data were collected with the Physics Toolbox Sensor Suite running on a smartphone. The app recorded three orthogonal magnetic-field components, $B_x(\mu\mathrm{T})$, $B_y(\mu\mathrm{T})$, and $B_z(\mu\mathrm{T})$, along with timestamps. The exported CSV file was opened in a data-analysis environment (Python/pandas). The time column was converted to a datetime format, and a subset of the record corresponding to the interval 17{:}33{:}00--17{:}40{:}00 (local time on the 4th) was selected for analysis. Time-series plots of $B_x$, $B_y$, $B_z$ (nT) and $\lvert \mathbf{B} \rvert$ (nT) versus time were generated for this interval to examine short-term variations in the local geomagnetic field. No additional filtering or smoothing was applied.

We applied a paired t-test on $\Delta_{\mathrm{1min}}$ to test the null hypothesis of zero mean difference, computed a 95~percent confidence interval for the mean smartphone--G-857 difference, calculated the standard deviation of $\Delta_{\mathrm{1min}}$ as a measure of scatter, and evaluated the Pearson correlation coefficient between the smartphone magnetometer and the G-857 magnetometer to assess coherence in short-timescale variability. Where sample sizes would permit, this framework could be extended to nonparametric tests such as the Wilcoxon signed-rank test and to comparisons across pre-flare, peak, and post-flare sub-windows, but here we focus on the 14 paired minutes spanning 17{:}27--17{:}40~UT. 

\section{Results}
\subsection{Smartphone--Proton Magnetometer Agreement}
\begin{figure}
  \centering
  \includegraphics[width=\textwidth]{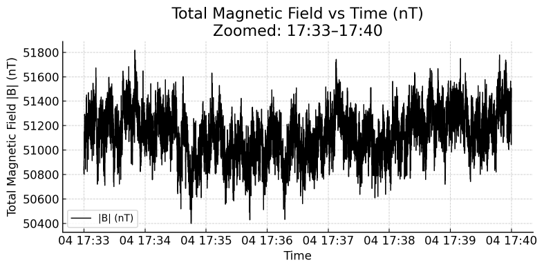}
  \caption{Unprocessed total magnetic-field observations with a smartphone magnetometer over UT 04{:}17{:}33 to 04{:}17{:}40.}
  \label{fig:raw}
\end{figure}

\begin{figure}
  \centering
  \includegraphics[width=\textwidth]{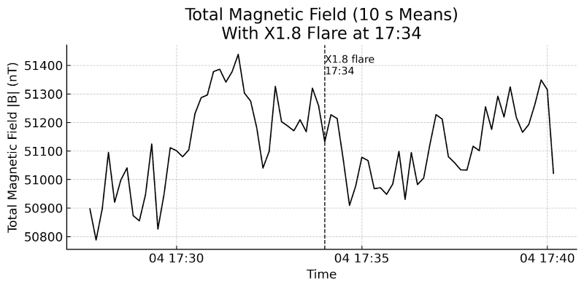}
  \caption{Processed total magnetic-field observations with a smartphone magnetometer, showing 10-second sample averages that substantially reduce the number of data points relative to the raw high-cadence data. These observations are over UT 04{:}17{:}33 to 04{:}17{:}40.}
  \label{fig:tensec}
\end{figure}

\begin{figure}
  \centering
  \includegraphics[width=\textwidth]{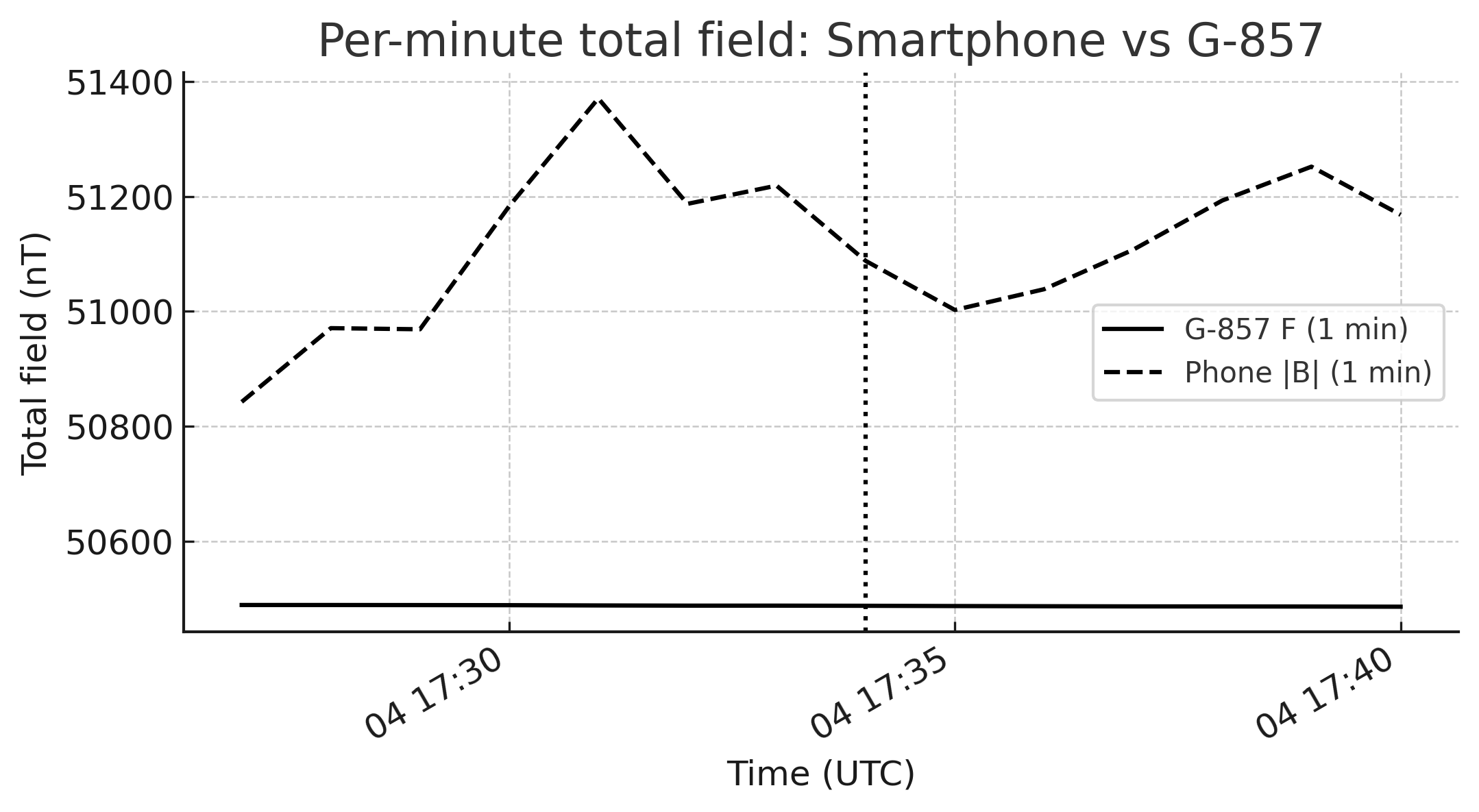}
  \caption{Processed total magnetic-field observations with a smartphone magnetometer and the G-857 magnetometer, showing one-minute averages over the broader flare-centered window.}
  \label{fig:paired}
\end{figure}

To quantify how well the smartphones tracked the reference instrument during the flare (Figure~\ref{fig:raw}--\ref{fig:paired}), we synchronized the raw smartphone (Figure~\ref{fig:raw}) and G-857 time series to a common universal-time base, computed per-minute averages for each instrument, and then restricted the analysis to a flare-centered window that maximized overlap while excluding minutes with missing or obviously bad data (Figure~\ref{fig:paired}). This yields a compact dataset in which each point represents contemporaneous, per-minute total-field estimates from both a smartphone magnetometer and a research-grade proton magnetometer under real mid-latitude space-weather conditions. Within the 17{:}27--17{:}40~UT interval surrounding the 4 November 2025 X1.8 flare, fourteen one-minute epochs contained valid, paired per-minute averages from both the smartphone magnetometer and the Geometrics G-857. The per-minute statistics are as follows: the number of paired minutes is 14, the mean smartphone field is 51{,}113.63~nT, and the mean G-857 field is 50{,}487.33~nT. The mean difference (phone minus G-857) is therefore $+626.30$~nT, with a standard deviation of 138.95~nT. The 95~\% confidence interval for the mean difference extends from 546.07 to 706.52~nT.

A paired t-test very strongly rejects the null hypothesis that the mean smartphone--G-857 difference is zero. The test yields a t statistic of 16.87 with a p value of approximately $3.2 \times 10^{-10}$, indicating an extremely statistically significant systematic offset, which can be clearly observed in Figure~\ref{fig:paired}. Thus, over this flare-centered window, the smartphones report total-field magnitudes that are, on average, approximately 626~nT higher than the research-grade proton magnetometer, with the mean offset constrained to within roughly $\pm 80$~nT at the 95~\% confidence level. From a geomagnetic perspective, this offset is large: it exceeds expected flare-time magnetic crochet amplitudes or similar rapid ionospheric signatures at mid-latitudes by one to two orders of magnitude, which are typically on the order of a few to a few tens of nanotesla (Curto et al., 1994, 2009; Curto, 2020). Even if such a crochet were present during the event, its signal would be fully submerged within the smartphone’s absolute calibration error.

\subsection{Coherence of Short-Timescale Variability}
Beyond the absolute offset, we assessed whether minute-scale fluctuations in the smartphone record track genuine geomagnetic variations captured by the G-857. The Pearson correlation between the per-minute smartphone and G-857 series over the 14 paired epochs is approximately $r \approx -0.38$, which is weak and negative. This indicates that when the G-857 total field increases slightly from one minute to the next, the smartphone often moves in the opposite direction or independently. The lack of coherence implies that short-timescale structure in the smartphone signal is dominated by instrumental noise, orientation effects, and local environmental contamination rather than by geophysical changes. Time-series visualization supports this interpretation: the per-minute G-857 series shows modest, smooth variability across 17{:}27--17{:}40~UT, consistent with quiet to mildly disturbed mid-latitude dayside conditions, whereas the smartphone series exhibits broad, erratic oscillations whose amplitude is comparable to the $\sim 600$~nT offset and bears no systematic relationship to the G-857 curve. The flare peak at 17{:}34~UT, clearly delineated in the GOES-18 soft X-ray profile, is not associated with any distinctive, flare-synchronous feature in the smartphone record.

Taken together, the smartphone--G-857 offset is 626.30~nT with a 95~\% confidence interval of 546.07--706.52~nT, and a paired t-test rejects the null hypothesis of zero mean difference ($t = 16.87$, $p \approx 3.2 \times 10^{-10}$). The Pearson correlation between the per-minute smartphone and G-857 series over the 14 paired epochs is $r \approx -0.38$, indicating weak, negative association. Even after 10-second and one-minute averaging, the smartphone time series retains residual variability on the order of several hundred nanotesla that does not track the smoother, low-amplitude variations present in the G-857 record.

\section{Discussion}
Our comparative analysis of smartphone magnetometer data and research-grade proton magnetometry during the 4 November 2025 X1.8 solar flare occupies a nexus of planetary magnetism, space weather, citizen science, and STEM education. Planetary magnetic fields arise from self-excited fluid dynamos in electrically conducting interiors. At Earth, the resulting geomagnetic field is continually modulated by ionospheric and magnetospheric current systems driven by the solar wind and transient events such as flares and CMEs (de Pater \& Lissauer, 2015; Stevenson, 2010). Representing this field with spherical-harmonic models and separating internal and external sources from global time series has been central to understanding secular variation, geomagnetic jerks, and disturbance fields (Kono, 2007). Operationally, networks such as INTERMAGNET and the USGS Geomagnetism Program maintain high-quality, absolute magnetometer records that resolve variations at the level of a few nanotesla, enabling both scientific analysis and space-weather services (INTERMAGNET, 2024; U.S. Geological Survey, 2025). Our campus Geometrics G-857 proton-precession magnetometer, co-located with smartphones and cross-checked against nearby INTERMAGNET-quality stations, effectively ``tapped into'' this global infrastructure at the local scale.

\subsection{Smartphone Magnetometry: Data Quality, Limitations, and Value}
Our quantitative results and qualitative comparisons point to a clear conclusion about smartphone use in geomagnetic monitoring. A central empirical finding is the stark discrepancy between smartphone magnetometer readings and the research-grade G-857 measurements. Across fourteen one-minute epochs spanning 17{:}27--17{:}41~UT, the smartphone’s reported total field was consistently higher than the G-857’s absolute value by an average of $\approx +626$~nT, with a narrow 95~\% confidence interval of roughly $+546$ to $+707$~nT and a highly significant mean difference (paired $t \approx 16.9$, $p \approx 3\times10^{-10}$). In practical terms, a $\sim 600$~nT bias is enormous: it represents $>1\%$ of Earth’s total field at our latitude ($\approx 50{,}000$~nT) and exceeds by one to two orders of magnitude the flare-time perturbations ($\sim 1$--10~nT) one might hope to detect (de Pater \& Lissauer, 2015; Curto, 2020). The per-minute correlation between smartphone and G-857 readings is weak and negative ($r \approx -0.38$), indicating that fluctuations in the smartphone record do not track legitimate short-timescale geomagnetic variability but instead reflect sensor noise, orientation-dependent biases, and local environmental contamination. Time-series visualization reinforces this interpretation: the G-857 record displays stable, gently varying behavior across the flare window, consistent with quiet mid-latitude dayside conditions, whereas the smartphone trace oscillates broadly and idiosyncratically---including through the 17{:}34~UT flare peak---bearing no resemblance to the proton magnetometer’s curve. Even after applying 10-second and one-minute averaging to suppress high-frequency noise, the smartphone signal remains dominated by absolute calibration errors on the order of hundreds of nanotesla, orientation effects that are difficult to characterize in field settings, and perturbations from nearby ferromagnetic materials and electronics. These effects collectively overwhelm the small-amplitude natural variations---on the order of a few to a few tens of nanotesla---that are of primary interest in space-weather studies such as crochet detection, storm-time variations, or quiet-time secular-field tracking. In contrast, the G-857, a research-grade proton magnetometer, delivers stable, absolute measurements of $F(t)$ that are directly comparable to regional observatories and suitable for detecting subtle geomagnetic variations when present. Under the observational conditions of this study, therefore, smartphone magnetometers are not suitable for quantitative geomagnetic monitoring or flare-time signature detection at mid-latitudes and cannot substitute for absolute total-field instruments in research or operational monitoring.

These findings dovetail with broader concerns about data quality in citizen science, where ``quality'' encompasses accuracy, precision, stability, representativeness, and transparency, and different stakeholders emphasize different facets (Balázs et al., 2021). Smartphone magnetometers---typically miniaturized anisotropic magnetoresistive or Hall-effect sensors---are designed primarily for orientation support, not precision scalar magnetometry. They are particularly susceptible to hard-iron and soft-iron effects, device-to-device heterogeneity, and local environmental interference. Our $\approx 600$~nT offset is consistent with uncorrected hard-iron bias compounded by orientation and environmental effects; the uncorrelated minute-scale variability reflects a noise floor far above the subtle geomagnetic signatures targeted in space-weather work. By contrast, proton-precession and fluxgate instruments at observatories routinely achieve absolute accuracies of a few nanotesla, resolve sub-nT variations, and operate within stringent calibration and quality-control regimes (INTERMAGNET, 2024; U.S. Geological Survey, 2025). In the terms articulated by Balázs et al. (2021), our explicit negative result is itself valuable: it clarifies that in this domain, smartphones cannot deliver observatory-quality magnetometry and should not be over-interpreted as such.

Crucially, this does not render smartphones useless. On the contrary, our results reinforce a growing consensus that smartphones are exceptionally powerful as pedagogical and citizen-science adjuncts, even as they fall short as primary research instruments. In our campaign, phones running Physics Toolbox Sensor Suite produced order-of-magnitude-reasonable total-field values (tens of microtesla) and responded qualitatively to gross environmental changes, allowing students to ``see'' that Earth’s magnetic field exists and varies slowly during the day (Vieyra \& Vieyra, 2019; de Pater \& Lissauer, 2015). Taking synchronized readings, comparing them with G-857 and observatory values, and discovering systematic offsets and noise naturally prompted discussion of calibration, sensor design, units, and error sources---core issues in scientific measurement. This experience aligns with a decade of physics and science education work that deploys smartphones as ``labs in your pocket'' for motion, forces, sound, light, and magnetism (Vieyra \& Vieyra, 2014; Vieyra et al., 2015; Vieyra, Vieyra, \& Macchia, 2017; Vieyra, 2018). Gamified and AR-enhanced tools such as Physics Toolbox Play and Magna-AR use smartphones to make abstract concepts tangible across formal and informal settings (Vieyra et al., 2020; Vieyra \& Vieyra, 2022; Vieyra et al., 2023), while LiDAR-based modes like the LiDAR Motion Visualizer map embodied movement onto position--time graphs and have been shown to improve students’ interpretation of motion graphs (Megowan-Romanowicz et al., 2023; O’Brien et al., 2023; Vieyra et al., 2024). Parallel AR work uses phones to visualize 3-D magnetic field structures and link stored magnetic energy to solar flares (Vieyra et al., 2024), and reliability studies in applied contexts---for example, smartphone inertial sensors used for dynamic balance assessment---report good to excellent intra- and inter-rater reliability under controlled protocols (Atalay \& Turan Kızıldoğan, 2025). Our flare-centered experiment extends this pattern into geomagnetism: smartphones are well-suited as contextual, qualitative sensors and as catalysts for metacognitive reflection on measurement and data quality, but their effective resolution, calibration stability, and noise characteristics remain insufficient for detecting genuine changes in Earth’s magnetic field under real mid-latitude observing conditions.

\subsection{Geomagnetic Flare Effects in Context}
The coordinated observations during the 4 November 2025 X1.8 flare provide a mid-latitude case study of solar flare effects (SFEs) on Earth’s magnetic field. Over the 17{:}27--17{:}41~UTC window, the G-857 showed smooth total-field magnitudes with only slight natural variation, and students’ manual reads at 17{:}30, 17{:}34, and 17{:}40~UTC closely matched the nearest USGS/INTERMAGNET observatory, which provides sub-nT resolution and absolute calibration (INTERMAGNET, 2024; U.S. Geological Survey, 2025). No abrupt, flare-synchronous ``magnetic crochet'' was discernible above baseline fluctuations once reading uncertainty was considered. This non-detection is consistent with SFE climatology: rapid crochets from enhanced dayside ionospheric currents occur preferentially under specific conditions (near local noon, favorable solar zenith angle, quiet background), and their amplitude and occurrence depend strongly on latitude, local time, and conductance (Curto et al., 1994; Curto \& Gaya-Piqué, 2009; Curto, 2020). Many X-class flares at mid-latitudes, especially away from noon, produce ground perturbations of only a few to a few tens of nanotesla---comparable to quiet-time variability that generally requires detrending and multi-station analysis to isolate (Shibata \& Magara, 2011; Benz, 2017).

Recent work at high latitudes and in cusp regions has documented SFEs reaching hundreds of nanotesla during intense events, highlighting the geographical selectivity of strong responses (Yamauchi et al., 2020; Rao et al., 2024), whereas classic surveys show mid-latitude crochets typically remain in the few--few-tens of nT range (Curto et al., 1994; Curto \& Gaya-Piqué, 2009; Curto, 2020). In this context, our magnitude-only, single-station analysis---showing smooth G-857 behavior through the flare and close agreement with the observatory series---makes the absence of a clear crochet unsurprising and pedagogically useful. Multi-source ``triangulation'' was essential: GOES-18 soft X-ray flux provided a precise radiative clock for the flare’s impulsive phase, while USGS/INTERMAGNET data offered a calibrated regional baseline (INTERMAGNET, 2024; U.S. Geological Survey, 2025). The agreement between campus and observatory records indicates that any genuine flare-time effect at our site lay below the detection threshold of simple magnitude reads. In SFE research, such cross-validation guards against false positives and ensures claimed crochets exceed normal diurnal variation and instrument noise (Curto, 2020; Curto et al., 2017; Meza et al., 2009; Selvakumaran et al., 2015). For students and citizen scientists, this framework underscores an important epistemic lesson: sometimes ``not seeing anything'' is the correct scientific result, strengthened---not weakened---by careful comparison with authoritative reference data.

\subsection{Citizen Science, STEM Education, and Geomagnetic Monitoring}
Framing our flare campaign within the citizen-science literature helps clarify its dual scientific and educational goals. Across environmental and ecological domains, citizen science has matured into a full life-cycle practice---from co-design and implementation to evaluation and long-term data management---while grappling with issues of engagement, data quality, and ethics (Fraisl et al., 2022; Frigerio et al., 2021; Balázs et al., 2021). In astronomy and geospace sciences, citizen projects now span classification of galaxies and light curves, optical transient discovery, and community-hosted instrumentation, often occupying niches poorly served by major observatories or automated algorithms (Marshall et al., 2015; Grandin et al., 2025). However, as Bonney et al. (2016) emphasize, robust scientific outputs do not automatically yield robust gains in participants’ understanding of science. Measurable conceptual and epistemic gains tend to arise when projects are intentionally designed to foreground scientific practices, uncertainty, and the nature of scientific knowledge.

Auroral displays have been documented for millennia---for example in the Bamboo Annals near Xi’an---and are now understood as signatures of intense solar--terrestrial coupling that usually remain confined to high latitudes except during strong geomagnetic storms (Usoskin et al., 2023; Blake et al., 2021; Grandin et al., 2024a). The 10 May 2024 ``Gannon Storm,'' the most powerful event since the 2003 Halloween storms, was driven by a sequence of CMEs from a magnetically complex sunspot group roughly 16 Earth diameters across, producing exceptionally efficient solar wind--magnetosphere coupling and global auroral activity (Greshko, 2024; Spogli et al., 2024; Kwak et al., 2024; Grandin et al., 2024a). Such superstorms threaten power grids, satellites, GNSS, HF radio, and aviation, with estimated economic impacts from tens of billions of dollars per day to trillions for Carrington-class events, though these estimates remain highly uncertain (Hapgood et al., 2021; Oughton et al., 2017; Eastwood et al., 2018; Grandin et al., 2024a). Historical benchmarks such as the 1859 Carrington storm and the near-miss CME of 23 July 2012 demonstrate that extreme events are both physically plausible and societally consequential, underscoring the need to characterize their evolution (Siscoe et al., 2006; Baker et al., 2013; Grandin et al., 2024a).

Ironically, the observational infrastructure is least adequate precisely when conditions are most extreme. During events like the Gannon Storm, the auroral oval expands into mid- and low-latitude regions with sparse optical and radar coverage, and even extensive networks like SuperDARN provide patchy mid-latitude sampling, with notable gaps across Europe, Russia, and the Southern Hemisphere (Johnsen, 2013; Nishitani et al., 2019; Kataoka et al., 2024; Grandin et al., 2024a). Auroral precipitation models such as OVATION Prime are calibrated mostly on moderate storms and tuned for $K_p \lesssim 8+$, while AI-based forecasts are trained on datasets dominated by typical conditions, so a $K_p = 9$ superstorm like the Gannon event pushed these tools beyond their validated regime, leading to mislocated boundaries and underestimates of auroral extent and intensity (Newell et al., 2014; Oliveira et al., 2021; Spogli et al., 2024; Grandin et al., 2024a). This mismatch between disturbance severity and model validity motivates complementary observing strategies that can document superstorms at scales and locations where conventional systems and models struggle.

One of the most important emerging complements is the global community of aurora observers equipped with high-sensitivity consumer and smartphone cameras. Over the past decade, coordinated analysis of citizen imagery has enabled the discovery and characterization of STEVE (Strong Thermal Emission Velocity Enhancement) and its distinctive green ``picket fence,'' revealing unique spectral and morphological properties that standard networks had not captured (MacDonald et al., 2018; Archer et al., 2019; Mende \& Turner, 2019; Mende et al., 2019; Martinis et al., 2021; Nishimura et al., 2023). Citizen images have also clarified the relationship between STEVE and stable auroral red arcs, revealed faint streak-like emissions below STEVE, and documented fragmented auroral patches and other fine-scale structures difficult to resolve with conventional imaging systems (Semeter et al., 2020; Whiter et al., 2021; Martinis et al., 2022). In addition, citizen photography played a central role in discovering and analyzing ``dunes,'' wave-like patterns in diffuse green aurora that probe mesospheric and lower-thermospheric dynamics (Palmroth et al., 2020; Grandin et al., 2021; He et al., 2023). Collectively, these studies show that carefully vetted citizen-science data can sample spatial and temporal scales, and geographic regions, that are undersampled by traditional auroral monitoring (MacDonald et al., 2018; Palmroth et al., 2020; Grandin et al., 2021; Grandin et al., 2024a).

Citizen-science platforms generalize this from isolated images to statistically rich, geospatially structured datasets. Aurorasaurus aggregates crowd-sourced sighting reports and social-media posts to generate near-real-time auroral visibility maps and provide independent checks on model-derived auroral boundaries (MacDonald et al., 2015; Case et al., 2016a; Kosar et al., 2018a, 2018b). Skywarden (Taivaanvahti), a Finnish-based, but globally used aurora catalogue, contains over 10{,}000 quality-controlled observations with associated images and metadata, supporting research on dunes, proton-injection--related red arcs, and other atmospheric optical phenomena (Karttunen, 2021; Gritsevich et al., 2014; Moilanen \& Gritsevich, 2022; Palmroth et al., 2020; Nishimura et al., 2022). Together with related initiatives, these platforms function as a distributed observing system in which thousands of volunteers act as geographically dispersed sensors whose time-stamped, geolocated reports can be cross-referenced with satellite and ground-based measurements (Ledvina et al., 2023; Grandin et al., 2024a).

The Gannon Storm is the first geomagnetic superstorm of its intensity to occur in an era when smartphones, low-light cameras, social media, and mature citizen-science infrastructures are all widely available, making it a natural test case for this distributed paradigm (Grandin et al., 2024a). The ARCTICS working group responded with a rapid online survey to document auroral visibility and technology disruptions on 10 May 2024, collecting hundreds of reports from more than 30 countries, including many at geomagnetic latitudes of $\sim 30^\circ$--$50^\circ$---well equatorward of typical auroral zones and empirical oval predictions---and frequent descriptions of intense red and pink emissions indicative of large fluxes of low-energy electrons (Grandin et al., 2024a). At the same time, uneven geographic coverage, variable timing precision, and heterogeneous descriptive language revealed the limitations of ad hoc reporting and highlighted the need for more standardized citizen-science protocols in future extreme events (Grandin et al., 2024a).

Our campus-based flare campaign was conceived as a small-scale test of how low-cost, citizen-centered approaches might complement such professional and citizen-science efforts. With a single G-857 magnetometer, ubiquitous smartphones, and access to GOES and USGS/INTERMAGNET data, a university team conducted a coordinated flare observation that intersected meaningfully with operational streams. Students engaged in professional-like workflows---timestamping against GOES, aligning multi-source time series, computing $\Delta F$, and interpreting uncertainty---consistent with research-based recommendations for authentic data practice in astronomy and physics education (Prather et al., 2004, 2009; Grandin et al., 2025). Tying local measurements to an INTERMAGNET observatory record and GOES X-ray timeline provided a scaffold for responsible participation: claims of a local flare effect were evaluated against authoritative context, and students ultimately did not claim a crochet, exemplifying structured skepticism.

The project also clarified the promise and limits of low-cost instrumentation. Experience with low-cost resistivity and seismic refraction systems shows that carefully designed devices can approach commercial-grade performance (Clark \& Page, 2011), suggesting that community fluxgate stations or personal magnetometer nodes could similarly extend magnetometer coverage (Curto, 2020; Grandin et al., 2025). In this vision, citizen volunteers host mid-quality magnetometers at schools and community sites, contributing continuous time series that, once calibrated against observatory records, would enhance spatial coverage of solar flare effects and storm-time disturbances, while smartphones act primarily as engagement tools that help participants explore local anomalies and understand why higher-end instruments are needed (Vieyra \& Vieyra, 2014; Vieyra et al., 2015). The HamSCI Personal Space Weather Station exemplifies this approach, with a low-cost, COTS-based magnetometer using magneto-inductive sensors ($\sim 3$~nT resolution at 1~Hz over $\pm 1.1\times10^6$~nT and $-40$ to $+85\,^\circ$C) that closely tracks research-grade instruments, and which we plan to deploy for ongoing monitoring and education (Regoli et al., 2018; Frissell et al., 2023; Kim et al., 2024).

Our intentionally simplified flare analysis points toward more advanced citizen-science campaigns aimed at formal crochet detection. Key requirements include multi-station coverage across critical local-time and latitude sectors to exploit SFE geometry, detrending and automated detection algorithms to isolate step-like signatures, and multi-parameter context (GNSS-TEC, VLF/HF propagation, ionosondes) to confirm radiative forcing and separate competing processes (Curto et al., 1994, 2017; Meza et al., 2009; Selvakumaran et al., 2015; Curto, 2020; Yamauchi et al., 2020; Rao et al., 2024; Grandin et al., 2025). Citizen scientists can contribute by hosting instruments, maintaining long time series, and assisting with data labeling and validation, provided that projects adopt explicit data-quality and reporting frameworks (Balázs et al., 2021; Fraisl et al., 2022).

Situating these efforts within the broader citizen-science literature emphasizes that they are as much about learning as about data. Across environmental and ecological domains, citizen science has evolved into a full research pipeline---from co-design to long-term data stewardship---while continually addressing engagement, data reliability, and ethics (Balázs et al., 2021; Frigerio et al., 2021; Fraisl et al., 2022). Analogous developments in astronomy and geospace include large-scale classification projects, community-hosted instrumentation, and local monitoring initiatives that fill niches beyond the reach of major observatories (Marshall et al., 2015; Grandin et al., 2025). Yet high-quality datasets do not automatically yield deeper public understanding; conceptual and epistemic gains emerge when projects explicitly foreground scientific practices, uncertainty, and the provisional nature of scientific claims (Bonney et al., 2016).

Our flare campaign was therefore deliberately framed as both research and pedagogy. By embedding explicit nature-of-science reflection in a content-rich investigation of solar flare effects, we followed Abd-El-Khalick’s (2001) recommendation that NOS instruction be integrated into authentic inquiry rather than isolated in methods courses. Students confronted questions about signal versus noise, instrument adequacy, and the meaning of null results, aligning with data-rich, place-based curricula such as Measuring the Earth with GPS and community-college astronomy projects that couple skyglow surveys, stargazing, and beach plastic investigations to disciplinary learning objectives (Kortz \& Smay, 2019; Prather et al., 2004, 2009; Hart et al., 2025). Participation in framing the research question, deploying instruments, and interpreting ambiguous data is consistent with evidence that short, well-structured research experiences can strengthen scientific identity and understanding (Abd-El-Khalick, 2001; Bonney et al., 2016; Prather et al., 2004; Hart et al., 2025).

By linking campus measurements to broader questions in planetary magnetism and geophysics---such as how magnetometry constrains interior dynamics, magnetic mineralogy, and serpentinization processes relevant to planetary habitability---we emphasized that space weather sits within a continuum connecting solar activity, planetary interiors, and technological society (de Pater \& Lissauer, 2015; Stevenson, 2010; Kono, 2007; Hart, 2024; Hart et al., 2023). The side-by-side deployment of smartphones, a research-grade G-857, and observatory data illustrated an instrumentation ladder from entry-level citizen tools to professional sensors, demystifying high-end science and empowering communities to begin their own measurements (Vieyra \& Vieyra, 2014; Vieyra et al., 2015; Bonney et al., 2016; Hart et al., 2025). In this sense, the Gannon Storm and our associated campaigns exemplify a broader transformation: extreme space-weather events are becoming opportunities not only to extend observational reach via global citizen-science infrastructures, but also to create NOS-informed learning environments that foster scientifically literate participants capable of contributing responsibly to space-weather and geophysical monitoring in an increasingly technology-dependent world (Ledvina et al., 2023; Grandin et al., 2024a, 2025).

\subsection{Lower-Cost Monitoring and Broader Participation in Earth and Space Science}
Our findings also connect to the broader theme of low-cost geophysics and the scalability of community monitoring. In near-surface geophysics, Clark and Page (2011) show that carefully designed, inexpensive resistivity and seismic refraction systems can produce subsurface images comparable to commercial equipment, with trade-offs focused on convenience and automation rather than core data quality. A homemade resistivity meter yielded results comparable to an AGI Sting R1 system, and a simple 12-channel seismograph produced depth estimates within approximately 0.4~m of a US\$15k commercial unit (Clark \& Page, 2011). These examples demonstrate that cost and quality are not inevitably coupled: with careful calibration, transparent methods, and realistic expectations, low-cost instruments can support valid geophysical inference (Clark \& Page, 2011).

A similar logic applies in geomagnetism and planetary science, where magnetometry and magnetic mineralogy are central to constraining interior processes and assessing planetary habitability. Spacecraft magnetometers operating at nanotesla sensitivity have been crucial for revealing planetary dynamos, remanent crustal fields, and their evolution (Stevenson, 2010; Kono, 2007; de Pater \& Lissauer, 2015). On Earth, detailed magnetic studies of variably serpentinized ultramafic rocks and extraterrestrial analogs link crustal magnetization, hydrothermal alteration, and potential energy sources for life (Hart, 2024; Hart, Cardace, \& Kennedy, 2023). In this context, understanding what magnetic signatures can and cannot be robustly extracted from a given sensor is a prerequisite for drawing reliable geophysical and astrobiological inferences. Our campus flare experiment is conceptually parallel: although focused on space-weather perturbations, it effectively estimates the practical resolution and bias of smartphone magnetometers relative to a research-grade scalar instrument, clarifying where phones sit on an ``instrumentation ladder'' that also includes DIY boards and observatory-grade sensors.

Under realistic mid-latitude observational conditions, smartphone magnetometers lack the precision, stability, and calibration required to detect genuine geomagnetic variations associated with solar flares or other ionospheric processes, whereas a research-grade proton magnetometer integrated into the global observatory network can provide stable, accurate measurements at the needed sensitivity. The smartphone data exhibited a systematic offset of $\approx +600$~nT relative to the G-857 and a weak, negative correlation with proton magnetometer readings, with noise levels far exceeding the expected amplitude of flare-time solar flare effects (SFEs) at our latitude. Our non-detection of a flare-synchronous crochet at both campus and observatory stations is consistent with established SFE climatology: many X-class flares do not produce clearly resolvable ground signatures at all locations, especially at mid-latitudes and away from local noon (Curto et al., 1994; Curto \& Gaya-Piqué, 2009; Curto, 2020). Thus, the ``negative'' smartphone result is not a failure but a clear empirical boundary on what phones can and cannot do.

This analysis feeds directly into the instrumentation ladder needed for scalable citizen science in geomagnetism. At the top, proton and fluxgate magnetometers in national networks (e.g., INTERMAGNET and USGS observatories) provide authoritative, high-precision reference data (INTERMAGNET, 2024; U.S. Geological Survey, 2025). Intermediate rungs might consist of purpose-built, moderately priced magnetometer kits deployed at schools or community centers, while the bottom tier comprises smartphones and ultra-low-cost sensors that act primarily as high-engagement entry points, supporting broad participation, qualitative exploration, and metacognitive reflection (Clark \& Page, 2011; Vieyra \& Vieyra, 2014; Vieyra et al., 2015; Grandin et al., 2025). Strategically combining these tiers---calibrating lower-tier data against higher-tier references---could dramatically increase spatial coverage for mapping Earth’s magnetic response to solar activity, particularly in under-instrumented regions, while simultaneously expanding educational reach.

Low-cost instrumentation beyond smartphones has real potential if appropriately engineered and deployed. The low-cost resistivity and seismic examples (Clark \& Page, 2011) show that carefully designed systems can approach commercial-grade data quality, and similar efforts in magnetometry---such as community fluxgate stations or personal magnetometer nodes---are already emerging (Curto, 2020; Grandin et al., 2025; Kim et al., 2024). One can envision a distributed network where citizen volunteers host mid-quality magnetometers at schools and community sites, contributing continuous data streams that, once calibrated against observatory records, enhance spatial coverage of SFEs and storm-time disturbances (Curto, 2020; Grandin et al., 2025). Smartphones in such a network would not serve as core sensors but as engagement tools, helping participants calibrate their intuition, explore local anomalies, and understand why higher-end instruments are necessary for subtle signals (Vieyra \& Vieyra, 2014; Vieyra et al., 2015).

The HamSCI Personal Space Weather Station provides a concrete template for this vision. This effort has developed a low-cost, commercial-off-the-shelf, easy-to-assemble magnetometer using basic circuits and magneto-inductive sensors that measure three-axis magnetic-field variations with $\sim 3$~nT resolution at 1~Hz over a $\pm 1.1\times10^6$~nT range and a $-40$ to $+85\,^\circ$C temperature span (Regoli et al., 2018; Kim et al., 2024). Comparative studies show that these systems closely match research-grade magnetometers under realistic conditions (Regoli et al., 2018; Kim et al., 2024). Future work in our setting will involve installing one of these HamSCI designs for continued monitoring and science-education use, thereby populating the intermediate rung of the instrumentation ladder and linking campus-scale measurements to broader community and professional networks (Regoli et al., 2018; Frissell et al., 2023; Kim et al., 2024; Grandin et al., 2025).

These results suggest a coherent strategy: smartphones and other ultra-low-cost devices remain invaluable for engagement and coarse context, but precision monitoring of flare-time perturbations and storm-time dynamics requires an integrated hierarchy of instruments---from research-grade observatories down through community magnetometer kits---with explicit calibration pathways and clearly defined roles for each tier (e.g., Clark \& Page, 2011; Curto, 2020; Hart, 2024; Hart et al., 2023; de Pater \& Lissauer, 2015; Stevenson, 2010; Kono, 2007; Vieyra \& Vieyra, 2014; Vieyra et al., 2015; Grandin et al., 2025).

\section{Conclusions}
Viewed holistically, our analyses indicate that under realistic mid-latitude observing conditions, smartphone magnetometers do not possess the precision, long-term stability, or calibration fidelity required to resolve genuine geomagnetic variations associated with solar flares or related ionospheric processes, whereas a research-grade proton magnetometer embedded in the global observatory network does meet the necessary sensitivity and stability thresholds (See further Curto et al., 1994; Curto \& Gaya-Piqué, 2009; Curto, 2020; Shibata \& Magara, 2011; Benz, 2017). Quantitatively, the phone data show a systematic offset of $\approx +600$~nT relative to the G-857, a weak negative correlation with proton readings, and noise levels far exceeding the expected amplitude of flare-time SFEs at our latitude, and our non-detection of a flare-synchronous crochet at both campus and observatory sites is entirely consistent with SFE climatology, in which many X-class flares fail to produce clearly resolvable ground signatures at mid-latitudes and away from local noon (Curto et al., 1994; Curto \& Gaya-Piqué, 2009; Curto, 2020; Shibata \& Magara, 2011; Benz, 2017). At the same time, side-by-side deployment of smartphones, the G-857, and INTERMAGNET/USGS reference data created a powerful educational and citizen-science environment in which participants engaged in authentic practices---time synchronization, unit conversion, residual construction, and critical assessment of noise and bias---while seeing concretely why high-precision instruments and explicit data-quality frameworks are indispensable (Prather et al., 2004, 2009; Balázs et al., 2021; Fraisl et al., 2022). Consistent with prior work on smartphone-based science learning, our results reinforce that phones and other low-cost sensors function best as ``labs in your pocket'' that support conceptual understanding, inquiry, metacognition, and reflection on the nature of science when embedded in carefully scaffolded activities that foreground both capabilities and limitations (e.g., Vieyra \& Vieyra, 2014, 2019, 2022; Vieyra et al., 2015, 2017, 2020, 2023, 2024; Atalay \& Turan Kızıldoğan, 2025; Megowan-Romanowicz et al., 2023; O’Brien et al., 2023; Abd-El-Khalick, 2001; Bonney et al., 2016; Hart et al., 2025). 

Building on this, our findings argue for a tiered instrumentation architecture in which research-grade observatories define the reference standard; intermediate-cost, purpose-built magnetometers (e.g., HamSCI Personal Space Weather Station units using magneto-inductive sensors with $\sim 3$~nT resolution at 1~Hz over a $\pm 1.1\times10^6$~nT range and $-40$ to $+85\,^\circ$C) extend spatial coverage; and smartphones plus ultra-low-cost devices provide high-engagement entry points in geophysical sciences with applications in Earth and space sciences, while robust SFE detection further demands multi-station local-time coverage, detrending and automated step-detection, and integration of GNSS-TEC, VLF/HF, and ionosonde context (See Clark \& Page, 2011; Meza et al., 2009; Selvakumaran et al., 2015; Curto et al., 2017; Yamauchi et al., 2020; Rao et al., 2024; Regoli et al., 2018; Frissell et al., 2023; Kim et al., 2024; Grandin et al., 2025). Within such a framework, citizen scientists and students can contribute meaningfully to long-baseline monitoring and event characterization---hosting instruments, maintaining time series, assisting with labeling and validation---while deepening their understanding of planetary magnetism, space weather, and interior--magnetosphere coupling (de Pater \& Lissauer, 2015; Stevenson, 2010; Kono, 2007; Hart et al., 2025). Ultimately, combining research-grade sensors with a distributed, well-scaffolded citizen network, informed by best practices in low-cost geophysics and participatory heliophysics, offers a synergistic route to a more densely sampled and educationally rich picture of Earth’s magnetic environment and its response to the dynamic Sun (Clark \& Page, 2011; Regoli et al., 2018; Frissell et al., 2023; Kim et al., 2024; Ledvina et al., 2023; Grandin et al., 2025).


\section{Declaration of Conflicts of Interest}
The authors declare no conflict of interest.

\section{Acknowledgments}
We thank NOAA’s National Centers for Environmental Information (NCEI) and the Space Weather Prediction Center (SWPC) for the GOES X-ray, proton, and magnetometer data products; the U.S. Geological Survey Geomagnetism Program and INTERMAGNET for observatory data; and the students and staff at the Community College of Rhode Island for installing and operating the campus Geometrics G-857 magnetometer. This work was supported by the National Science Foundation under Grant 2514197 and by the National Aeronautics and Space Administration through the NASA Rhode Island Space Grant award 80NSSC20M0053, subaward 00002512. Any opinions, findings, conclusions, or recommendations are those of the author and do not necessarily reflect the views of the National Science Foundation, NASA, NASA Rhode Island Space Grant, or any affiliated institutions. 

\section{Declaration of generative AI and AI-assisted technologies in the writing process}
During the preparation of this work, the author(s) have used generative AI tools for copy editing as AI-assisted improvements to human-generated text for readability and style, and to ensure that the text is free of errors in grammar, spelling, punctuation, and tone; these AI-assisted improvements may include wording and formatting changes. 

\section{Author Contributions}
Conceptualization and Lead, R.H.; Methodology, R.H.; Software, R.H.; Validation, R.H.; Formal analysis, R.H., L.M., E.S., S.K., I.F., G.S., Z.L., R.W., S.M., M.R., and B.G.; Investigation, R.H., L.M., E.S., S.K., I.F., G.S., Z.L., R.W., S.M., M.R., and B.G.; Resources, R.H.; Data curation, R.H.; Writing—original draft, R.H.; Writing—review and editing, R.H., L.M., E.S., S.K., I.F., G.S., Z.L., R.W., S.M., M.R., and B.G.; Visualization, R.H.; Supervision, R.H.; Project administration, R.H., L.M.; Funding acquisition, R.H. All authors have read and agreed to the published version of the manuscript.

\end{document}